\documentstyle[12pt,psfig]{article}
\begin{document}
\renewcommand{\thefootnote}{\fnsymbol{footnote}}

\vspace*{60pt}

\begin{flushleft}
{{\bf PHASE SEPARATION IN MODELS FOR MANGANITES: THEORETICAL ASPECTS 
AND COMPARISON WITH EXPERIMENTS}\footnote{To appear in proceedings 
of the workshop ``Physics of Manganites'', Michigan State University, 
July 26--29, 1998, eds. by T. A. Kaplan and S. D. Mahanti, Plenum 
Publishing Corporation. 
}}
\end{flushleft}

\vspace{36pt}

\begin{flushleft}
{\hspace{2.5cm} E. Dagotto, S. Yunoki, and A. Moreo\par}
\vspace{12pt}
{\hspace{2.5cm}National High Magnetic Field Lab and Department of Physics,\par}
{\hspace{2.5cm}Florida State University, Tallahassee, FL 32306, USA}

\end{flushleft}

\vspace{36pt}

\begin{flushleft}
{\bf INTRODUCTION}
\end{flushleft}

Colossal magnetoresistance in  metallic oxides such as
${\rm R_{1-x} X_x Mn O_3}$
$({\rm where~ R= La,}$
${\rm Pr,Nd;}$${\rm ~X=Sr,Ca,Ba,Pb})$
is attracting considerable
attention~\cite{jin} due to its potential technological applications.
A variety of experiments have revealed that  oxide
manganites have a rich phase diagram~\cite{phase} with regions of 
antiferromagnetic
(AF) and ferromagnetic (FM) order, as well as charge ordering, and
a peculiar insulating state above the FM critical temperature $T_c$.
Recently, $layered$ manganite compounds
${\rm La_{1.2} Sr_{1.8} Mn_2 O_7}$
have also been synthesized~\cite{moritomo} with properties similar to
those of their 3D counterparts.
Strong correlations are
important for transition-metal oxides, and, thus, theoretical guidance is
needed to understand the behavior of manganites and for the design of new
experiments.

The appearance of ferromagnetism at low temperatures can be explained using
the so-called
Double Exchange (DE) mechanism~\cite{zener,degennes}. However, the DE model
is incomplete to describe the entire phase diagram observed
experimentally. For instance, the
electron-phonon coupling may be crucial
to account for the insulating properties above
$T_c$~\cite{millis}. The presence of a Berry phase
at large Hund-coupling also challenges predictions
from the DE model~\cite{muller}. In a series of recent 
papers (Refs.~\cite{yunoki,yunoki_,yunoki2,yunoki3}) we have remarked
that another phenomena occurring in manganites
and not included in the DE description, namely the charge ordering
effect, may be contained in a more fundamental Kondo-like 
model.
More precisely, in those papers~\cite{yunoki,yunoki_,yunoki2} 
it was reported the presence
of {\it phase separation} (PS) between hole-undoped antiferromagnetic
and hole-rich ferromagnetic regions in the low
temperature phase diagram of the one-orbital FM Kondo model. In a
recent paper by the same authors a similar phenomenon was also observed
using two orbitals and Jahn-Teller phonons~\cite{yunoki3}. 
In this case the two phases
involved are both spin ferromagnetic and the $orbital$ degrees of
freedom are responsible for the phase separation.
Upon the inclusion of long-range
Coulombic repulsion,  charge ordering
in the form of non-trivial extended structures (such as stripes)
could be stabilized, similarly as
discussed for the cuprates~\cite{tj1,review,tranquada,tj2} but
now also including ferromagnetic domains.
The analysis carried out by our group suggests
that phenomena as rich as  observed
in the high-Tc superconductors may exist in
the manganites as well, and hopefully our
effort will induce further theoretical and
experimental work in this context.

The present paper should be considered as an informal
review of the present status of the computational studies of
models for manganites that reported the existence of phase
separation at low temperatures. It does $not$ pretend to be
a comprehensive article, and thus we encourage the readers
to consult the literature mentioned here to find out additional
references. The paper
is organized as follows: in the next section 
the results for the case of the one-orbital Kondo model are
reviewed, with emphasis on the phase diagram. Most of the calculations 
are performed with classical $\rm t_{2g}$-spins but 
results for quantum spins are also shown. In addition, models that
interpolate between Cu-oxides and Mn-oxides have also been studied,
as well as the influence on our conclusions
of Coulomb interactions beyond the on-site term.
In the following section results recently
reported for the case of the two-orbital model with Jahn-Teller phonons
are described. Once again the main emphasis is given to the phase
diagram. In both of those previous sections the main result is that
tendencies to ferromagnetism and phase-separation are in strong
competition in these models. Such a phenomenon appears so clearly in
all dimensions of interest and for such a wide range of models that
it leads us to believe that it may be relevant for experiments
in the context of manganites. A discussion of experimental literature
 that have
reported some sort of charge inhomogeneity compatible with phase
separation (after Coulomb interactions are added) are briefly described
towards the end of the paper. In the last section a summary is provided.

\begin{flushleft}
\vspace{24pt}
{\bf ONE ORBITAL FERROMAGNETIC KONDO MODEL}
\end{flushleft}

\begin{flushleft}
{\bf Results for Classical $t_{2g}$-Spins}
\end{flushleft}

The FM Kondo Hamiltonian~\cite{zener,furukawa} is defined as
$$
H = -t \sum_{\langle {\bf i j} \rangle \sigma} (c^\dagger_{{\bf i}\sigma} c_{{
\bf
j}\sigma} + h.c.) - J_H \sum_{{\bf i}\alpha \beta}
{ { c^{\dagger}_{{\bf i}\alpha} {\bf \sigma}_{\alpha \beta} c_{{\bf
i}\beta} }
\cdot{{\bf S}_{\bf i}}},
\eqno(1)
$$
\noindent where $c_{{\bf i}\sigma}$ are destruction
operators for one species of $e_g$-fermions at site ${\bf i}$
with spin $\sigma$,
and ${\bf S}_{\bf i}$ is the total
spin of the $t_{2g}$ electrons, assumed
localized.
The first term is the $e_g$ electron transfer between nearest-neighbor
Mn-ions,
$J_H>0$ is the Hund coupling, the number of sites is $L$,
and the rest of the notation is standard. The density is adjusted using
a chemical potential $\mu$.
In this section and Refs.~\cite{yunoki,yunoki_}
the spin ${\bf S}_{\bf i}$ is considered classical
(with $|{\bf S}_{\bf i}| = 1$),
unless otherwise stated.
Although models beyond Eq.(1) may be needed to fully understand
the manganites (notably those that include
lattice effects as studied later in this paper),
it is important to analyze in detail the properties of simple
Hamiltonians  to clarify if part  of the experimental rich phase
diagram can be accounted for using
purely electronic models.

To study  Eq.(1) in the $t_{2g}$-spin classical
limit a Monte Carlo (MC) technique was
used: first, the trace over the $e_g$-fermions
in the partition function was carried out
exactly diagonalizing the $2L \times 2L$ Hamiltonian of electrons
in the background of the spins \{${\bf S}_{\bf i}$\},
using library subroutines.
The fermionic trace is a positive function of the classical spins
and the resulting integration over the two angles per site
parametrizing the ${\bf S}_{\bf i}$ variables can be performed with
a standard MC algorithm without ``sign problems''.
In addition, part of the calculations of Refs.~\cite{yunoki,yunoki_} 
were also performed with the
Dynamical Mean-Field approximation (D=$\infty$)~\cite{furukawa}, the
Density-Matrix Renormalization Group (DMRG), and the Lanczos method.
Special care must be taken with
the boundary conditions (BC)~\cite{jose,zang}.

Our results
are summarized in
the  phase diagram
of Fig.1. In 1D (and also in 2D, not shown) and at low temperatures
clear indications of
(i) ferromagnetism, (ii) spin incommensurate (IC) correlations, and (iii)
phase separation were identified.
Results are also available in small 3D clusters and
qualitatively they agree with those in Fig.1. The same occurs 
working at  $\rm D=\infty$ (see Refs.~\cite{yunoki,yunoki_}).
In 1D we also obtained results with quantum $t_{2g}$-spins
S=3/2 (shown below) which are in good
agreement with Fig.1.

\begin{figure}[htb]
\vspace{-1.0cm}
\centerline{\psfig{figure=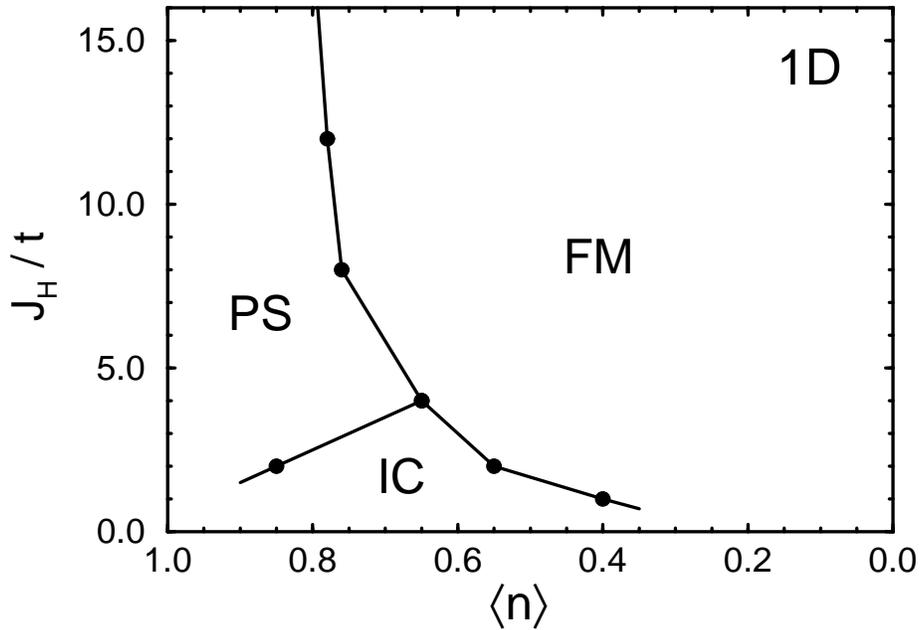,width=14.0cm,angle=-90}}
\vspace{0.2cm}
\caption{
Phase diagram of the FM Kondo model reported 
in Refs.~[8, 9] obtained
using numerical techniques.
FM, IC, and PS denote regimes with FM correlations,
spin incommensurate correlations,  and with
phase separation between undoped AF
and hole-rich FM regions, respectively.
The results shown correspond to one dimension and they were 
obtained with MC simulations
at  $T=t/75$ using chains with
$L$=20, 30 and 40 sites. This result is taken from 
Refs.~[8, 9].
}
\end{figure}

The boundaries of the FM region of the phase diagram
were found evaluating the spin-spin correlation
defined as $S({\bf q}) =
(1/L) \sum_{\bf j,m} e^{i {{\bf (j-m)}\cdot{\bf q}}}
\langle {{{\bf S}_{\bf j}}\cdot{{\bf S}_{\bf m}} }
   \rangle$. Fig.2
shows $S({\bf q})$ at zero momentum
vs. $T/t$ for typical examples in 1D and 2D.
The rapid increase of the spin correlations as $T$ is reduced and
as the lattice size grows clearly points towards the existence of
ferromagnetism in the system~\cite{foot3}.
It is natural to assume that
the driving force for FM in this context is the DE
mechanism.
Repeating this procedure for a variety of couplings and densities,
the robust region of FM shown in Fig.1 was determined.
In the small $J_H/t$ region IC correlations were
observed monitoring
$S({\bf q})$~\cite{inoue}.
Both in 1D and 2D there is one
dominant peak which moves away from  the
AF location  at $\langle n \rangle =1$ towards zero momentum as $\langle
n \rangle$ decreases. In the 2D clusters
the peak moves from
$(\pi,\pi)$ towards $(\pi,0)$ and $(0,\pi)$, rather than along
the main diagonal.
Note that our computational study predicts IC correlations only
in the small and intermediate
$J_H/t$ regime.

\begin{figure}[htb]
\vspace{-0.5cm}
\centerline{\psfig{figure=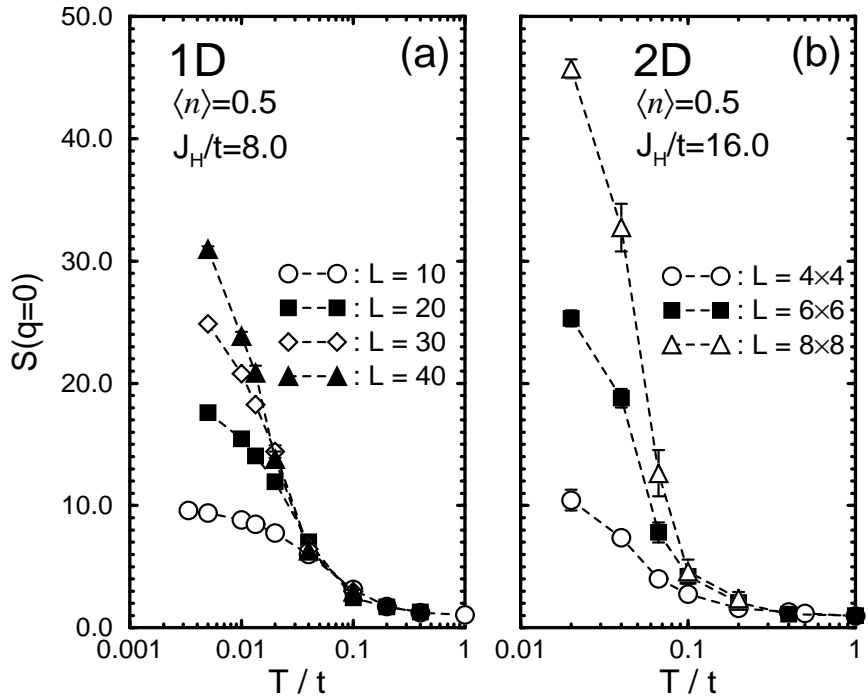,width=14.0cm,angle=-90}}
\vspace{0.2cm}
\caption{
Spin-spin correlations of the classical spins
 at zero momentum $S({\bf q}=0)$
vs temperature $T$ (in units of $t$).
MC results for several lattice
sizes are shown on (a) chains and (b) 2D clusters.
The density and coupling are shown.
In (a) closed shell BC are used i.e.
periodic BC for $L=10$ and $30$ and antiperiodic
BC for $L=20$ and $40$. In (b) open
BC are used. Results taken from Ref.~[8].
}
\end{figure}

The main result of Refs.~\cite{yunoki,yunoki_} is contained in
Fig.3 where the computational evidence for the existence of
phase separation in dimensions 1, 2, and $\infty$
is given. The presence of a
discontinuity in $\langle n \rangle$ vs $\mu$  shows that some
electronic densities can not be stabilized by tuning the chemical
potential~\cite{also}.
If the system is
nominally prepared with such density it will spontaneously
separate into two regions with the densities corresponding to the
extremes of the discontinuities of Fig.3.
By analyzing these extremes
the properties of the two domains can be studied.
One region is undoped ($\langle n \rangle = 1$)
with strong AF correlations, while the other contains all the holes
and the spin-spin correlations between the classical spins
are FM (see the inset
of Fig.3a. The results are similar in D=2 and infinite).
This is natural since holes optimize their
kinetic energy in a FM background. On the other hand, at
$\langle n \rangle =1$ the DE mechanism is not operative: if the electrons
fully align their spins they cannot move in the conduction band
due to the Pauli principle. Then, energetically
an AF pattern is formed.
As $J_H$ grows, the jump in Fig.3 is reduced and it tends to
disappear in the $J_H = \infty$ limit.

Experimentally, PS may be detectable using neutron diffraction
techniques, since $S(q)$ should present a
two peak structure, one located at the AF position and the other at zero
momentum. Since this also occurs in a canted ferromagnetic state care
must be taken in the analysis of the experimental data.
Another alternative is that Coulombic forces
prevent the macroscopic accumulation of charge intrinsic of a PS
regime. Thus, microscopic hole-rich domains may
develop in the manganites, perhaps in the form of stripes.

\begin{figure}[htb]
\vspace{0.20cm}
\centerline{\psfig{figure=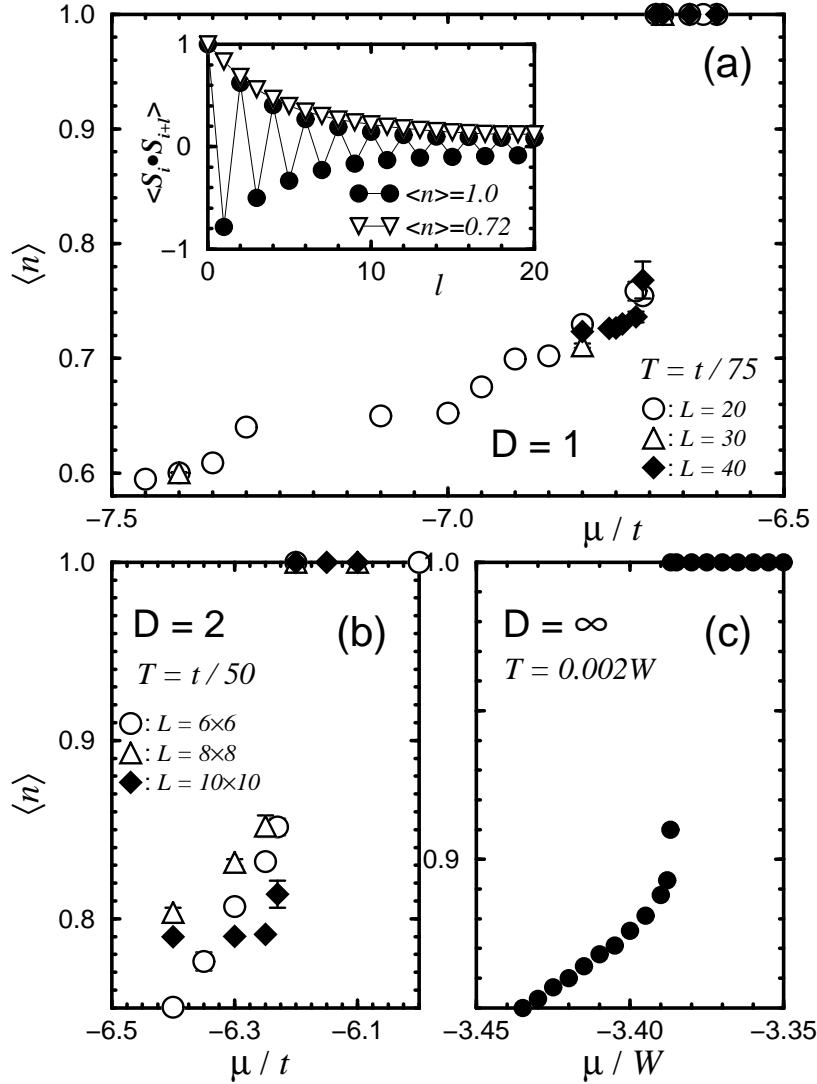,width=12.0cm,angle=-0}}
\vspace{0.2cm}
\caption{
Electronic density $\langle n \rangle$ vs chemical potential $\mu$
in (a) D=1, (b) D=2, and (c) ${\rm D=\infty}$ clusters.
Temperatures are indicated.
The coupling is $J_H/t=8.0$ in (a) and (b) and $J_H/W = 4.0$ 
in (c) (for the definition of $W$ see Ref.~[8]).
PBC were used
both in D=1 and 2. The discontinuities shown in the figures are indicative
of PS. In (a) the inset contains the spin-spin correlation in real space
at densities 1.0 and 0.72
showing that indeed PS occurs between AF and FM regions.
The results are taken from Refs.~[8, 9] 
where more details can be found.
}
\vspace{-0.2cm}
\end{figure}

Although the phase diagram of Fig.1 has PS
close to half-filling, actually this phenomenon
also occurs at  $\langle n \rangle  \sim 0$
if an extra direct AF exchange interaction between the
localized spins is included~\cite{yunoki2}. 
This coupling may be originated in a small hopping amplitude
for the $t_{2g}$ electrons. At $\langle
n \rangle =0$,
model Eq.(1) supplemented by a Heisenberg  coupling $J'/t$
 among the localized spins
 produces an AF phase, as in experiments,
 which upon electron doping
induces a
competition between AF (with no $e_g$-electrons)
and FM electron-rich regions, similarly as for $\langle n \rangle =1$
but replacing holes
by electrons.
Thus, PS or charge ordering
could exist in manganites both at large and small fermionic densities.
A careful study of the influence of the $J'$ coupling on
the phase diagram of the 1-orbital Kondo model has been recently
presented in Ref.~\cite{yunoki2}

For completeness,
upper bounds on  the 3D  critical temperature $T_c$
were also provided in Refs.~\cite{yunoki,yunoki_}. Using
MC simulations in principle it is possible to calculate
$T_c$  accurately. However,
the algorithm used in our studies prevented us from
studying clusters larger than $6^3$ even at $J_H=\infty$.
In spite of this limitation,
monitoring the spin-spin correlations in real space
allows us to judge at  what temperature  $T^*$
the correlation length reaches the boundary of
the $6^3$ cluster. Since the bulk $T_c$ is smaller than $T^*$, this
give us  upper bounds for the critical temperature.
Following this procedure at  $\langle n \rangle = 0.5$,
it was found that for $T \sim 0.1t$ robust
correlations reach the boundary, while for $T \ge 0.12t$ the correlation
is short-ranged. Thus, at this density we estimate that
$T_c < 0.12t$.
Since
results for the $e_g$ electrons bandwidth range
from $BW \sim 1~eV$~\cite{bandwidth} to $BW \sim 4 eV$~\cite{sarma},
producing a hopping $t = BW/12$ between $0.08$ and $0.33~eV$, then
the estimations for the critical temperature of Refs.~\cite{yunoki,yunoki_}
range  roughly  between
$T_c \sim 100~K$ and $400~K$.
This is within the experimental range and 
in agreement with other
results~\cite{furukawa,high}.
Then, in principle purely electronic models
can account for $T_c$.

\begin{flushleft}
\noindent{\bf Quantum {$\rm t_{2g}$}-Spins: Phase Diagram}
\end{flushleft}

In the previous section we have used classical localized spins to 
simplify the technical aspects of the computational study and since
the realistic values of those spins is large (3/2). However,
it is interesting to study the phase diagram for the case
of quantum $\rm t_{2g}$-spins at least  in one special case, in order to verify
that indeed the use of classical spins is a good approximation. Using
the Lanczos and DMRG techniques, the phase diagram for the cases of
$\rm t_{2g}$-spins $S=3/2$ and $1/2$ was reported in 
Refs.~\cite{yunoki,yunoki_} for
the case of one dimension, and one of them 
is shown in Fig.~4. The techniques used work in the canonical ensemble,
instead of grand-canonical as the MC simulation of the previous section. 
The presence of phase separation is investigated here using the
compressibility (which if it becomes negative signals the instability of
the density under study). Fig.~4 shows that the three regimes found
before (namely PS, FM, and IC) are also observed in the quantum case.
The results for $S=3/2$ are even quantitatively similar to those found
with the MC technique of the previous section, while those for $S=1/2$
are only qualitatively similar. The overall conclusion is that the
use of classical $\rm t_{2g}$-spins in the Ferromagnetic Kondo Model
was found to be a good approximation, and no important differences
in the phase diagram have
been observed (at least in this one-dimensional example) when quantum
spins are used. Studies in dimensions larger than one would be too involved
using quantum localized spins, and thus the similarities between 
the classical and quantum cases found here help us to investigate 
those more realistic dimensions.

\begin{figure}[htb]
\vspace{-0.5cm}
\centerline{\psfig{figure=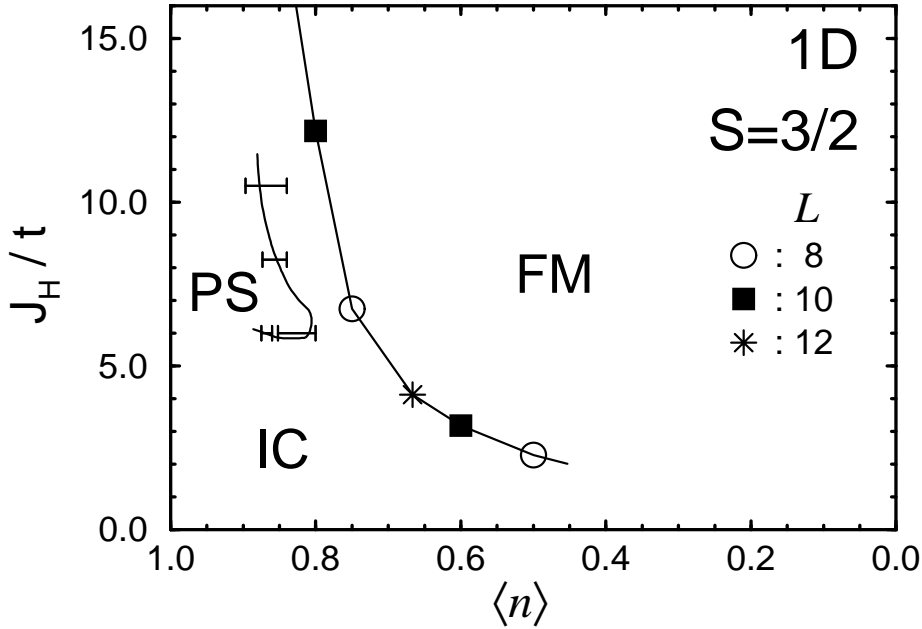,width=14.0cm,angle=-90}}
\vspace{0.2cm}
\caption{
Phase diagram of the Ferromagnetic Kondo model using
quantum $\rm t_{2g}$-spins. The results correspond to a localized
spin of value $S=3/2$ (normalized as $|S|=1$, see Ref.~[9]). 
The notation is the same as in Fig.1. 
The techniques used to
obtain the results are Lanczos and DMRG.
The result is taken from Ref.~[9].
}
\end{figure}

\begin{flushleft}
{\bf Models for Other Transition-Metal Oxides}
\end{flushleft}

In the previous sections we have found that models for manganites
analyzed with reliable unbiased techniques provide a phase
diagram where there are two main phases in competition: a ferromagnetic
regime and a ``phase-separated'' regime (actually the latter is not a
phase but a region of unstable densities). Such a clear result suggest
that the phenomenon maybe more general than expected. Actually recent
studies of models for transition-metal oxides in general~\cite{jose}, 
in the limit
of a large Hund-coupling, revealed phase diagrams where indeed FM and PS
dominated. Let us consider the evolution of such a phase diagram as
we move in the periodic table of elements  from copper, to nickel, etc,
eventually arriving to manganese. The study of Ref.~\cite{jose} was 
carried out
in one-dimension for technical reasons, but nevertheless the tendencies
observed are strong enough to be valid in other dimensions as well.
For the case of copper the model used was the plain ``t-J'' model, namely
a model where we have spins of value $S=1/2$ and holes which carry no
spin. The phase diagram in this case was known for a long time. It contains
no indications of ferromagnetism (at least no ``clear'' indications), and
there is phase separation only at large J/t. However, the study of 
Ref.~\cite{jose} for Ni-oxides
using a t-J-like model with ``spins'' of value $S=1$ and ``holes'' 
with spin $S=1/2$ (since they correspond to
the removal of one electron of a Ni-site) 
revealed that both FM and PS are clearly
enhanced in the phase diagram with respect to the case of Cu-oxides.  
The technical details can be found in Ref.~\cite{jose}.
As the spin grows these tendencies become stronger and actually for the
case of the Mn-oxides only a small window separates the fully polarized
FM and PS regimes. This result suggests that the phenomenon is very general
and should be searched for in other transition-metal oxides as well.

\begin{flushleft}
{\bf Influence of Coulomb Interactions}
\end{flushleft}

The accumulation of charge that occurs in the phase separation regime
described in the previous section is not stable once Coulombic
interactions beyond the on-site term are added to the problem.
It is expected that in the presence of these long-range interactions
the ground state will have clusters of one phase embedded into the
other (droplet regime). However, a computational study including
these Coulomb interactions is very difficult. For instance, the analysis
of model Eq.(1) would require a Hubbard-Stratonovich decomposition of the
fermionic interaction which not only introduces more degrees of freedom,
but in addition the infamous ``sign problem'' which will
prevent the study from being carried out at low enough temperatures. Then,
only techniques such as Lanczos or DMRG can handle properly this type
of models. Recently, the analysis including Coulombic terms
 started by using the ``minimum''
possible Hamiltonian, namely the 1-orbital  Ferromagnetic Kondo Model with
localized spin-1/2 and with the addition of an on-site Coulomb term and
a nearest-neighbor (NN) density-density $V$-coupling that penalizes the 
accumulation of charge in two NN sites~\cite{andre}. 
The results are still being
analyzed so here only a qualitative summary will be presented.

The conclusions of the study of Malvezzi et al~\cite{andre}. 
thus far are the
following: (1) the on-site coupling $U$ apparently does not affect
qualitatively the results found at $U=0$ as long as the Hund-coupling 
is large; (2) in the regime of phase separation near $\langle n \rangle =1$ 
the addition of a $V$-term produces results compatible with the cluster
formation anticipated in the previous paragraph. Actually even at finite
$V$ in a regime where the compressibility shows that phase separation
no longer exists (i.e. all densities are stable), clear remnants of such
a PS regime are found. For instance, at $V=0$ the spin structure factor
$S(q)$ has robust peaks at both $q=0$ and $\pi$. Introducing $V$, these
peaks still exist but they move slightly from the $0$ and $\pi$ positions
forming incommensurate structures; (3) in the regime where at $V=0$ there
is ferromagnetism induced by a double-exchange process, the inclusion of
a nonzero coupling $V$ transforms the ground state into a FM 
charge-density-wave (CDW). In this phase the spins are aligned but the
charge is not uniformly distributed.
The overall phase diagram is rather complicated.
We are actively working in this context to analyze the qualitative aspects
of such a result. But it is already clear that
charge-ordering tendencies are
observed at all densities upon introducing $V$, in agreement with 
experiments. More detailed results will be presented soon.

\begin{flushleft}
\vspace{24pt}
{\bf TWO ORBITALS AND JAHN-TELLER PHONONS}
\end{flushleft}

\begin{flushleft}
{\bf Phase Diagram}
\end{flushleft}

In spite of the rich phase diagram observed in the study of the
1-orbital Ferromagnetic Kondo Model in the previous section
and its similarities with 
experiments, there are aspects of the
real manganites that require a more sophisticated approach.
For instance, 
dynamical Jahn-Teller (JT) distortions are claimed to be  
very important~\cite{millis},
and a proper description of the
recently observed orbital
order~\cite{mura} obviously needs at least two orbitals.
Such a  multi-orbital model with JT phonons is  
nontrivial, and thus far it has been studied 
only using the dynamical mean-field approximation~\cite{millis2}.
The previous experience with the 1-orbital case suggests that 
 a computational analysis is actually
 crucial to understand its properties.
In addition, it is conceptually interesting to analyze whether 
the PS described before~\cite{yunoki,yunoki_,yunoki2} exists also 
in multi-orbital models.

The first
computational
study of a 2-orbital model for manganites including 
JT phonons was reported recently by the authors in Ref.~\cite{yunoki3}
and here a summary of the main conclusions will be discussed. 
The results show a rich phase diagram including
a novel regime of PS induced by the $orbital$, rather
than the spin, degrees of freedom (DOF). 
The Hamiltonian used in that study had three contributions
$H_{KJT} = H_K + H_{JT} + H_{AF}$. The first term is 
$$
H_K = -\sum_{{\bf \langle ij \rangle}\sigma a b} t_{ab}
(c^\dagger_{{\bf i} a \sigma} c_{{\bf j} b \sigma} + h.c.)
- J_H \sum_{{\bf i}a \alpha \beta} { {{\bf S}_{\bf i}}\cdot{
c^\dagger_{{\bf i}a \alpha} {\bf \sigma}_{\alpha \beta} c_{{\bf i}a \beta}    
 } },
\eqno{(2)}
$$
\noindent where ${\bf \langle ij \rangle}$ denotes nearest-neighbor
lattice sites,
$J_H > 0$ is the Hund coupling, the hopping amplitudes $t_{ab}$
are described in Ref.~\cite{hopping}, 
$a,b=1,2$ are the two $e_g$-orbitals, 
the $t_{2g}$ spins ${\bf S}_{\bf
i}$ are assumed to be classical 
(with $|{\bf S}_{\bf i}| = 1$)
since their actual value in $\rm Mn$-oxides (3/2) is large~\cite{approx},
and the rest of the notation is standard. 
None of the results
described below depends crucially on the set $\{ t_{ab} \}$
 selected~\cite{hopping}. The energy units are
chosen such that $t_{11}=1$ in the $x$-direction. In addition,
since $J_H$ is large in
the real manganites, here it will be  fixed to $8$ (largest
scale in the problem) 
unless otherwise stated. The $e_g$-density
$\langle n \rangle$ is adjusted using
a chemical potential $\mu$.

The coupling with JT-phonons is through~\cite{millis,kanamori}
$$
H_{JT} = \lambda \sum_{{\bf i} a b \sigma} c^{\dagger}_{{\bf i} a
\sigma}
Q^{ab}_{\bf i} c_{{\bf i} b
\sigma}
+ {{1}\over{2}} \sum_{\bf i} ( {Q^{(2)}}^2_{\bf i} + {Q^{(3)}}^2_{\bf i}),
\eqno{(3)}
$$
\noindent where $Q^{11}_{\bf i} = -Q^{22}_{\bf i} = Q^{(3)}_{\bf i}$, and
$Q^{12}_{\bf i} = Q^{21}_{\bf i} = Q^{(2)}_{\bf i}$. These phonons are assumed
to be classical, which substantially simplifies the computational study.
This is a reasonable first approximation towards the determination of
the phase diagram of the $H_{KJT}$ model.
Finally, a small coupling between the $t_{2g}$-spins is needed to 
account for the AF character of the real materials even
when all $\rm La$ is replaced by $\rm Ca$ or $\rm Sr$ 
(fully doped manganites). 
This classical
Heisenberg term is
$H_{AF} = 
J' \sum_{\bf \langle ij \rangle} {{{\bf S}_{\bf i}}\cdot{{\bf S}_{\bf j}}},
$
where $J'$ is fixed to $0.05$
throughout the paper, a value compatible with experiments~\cite{perring}. 
To study $H_{KJT}$
a Monte Carlo (MC) technique similar to that employed in
Refs.~\cite{yunoki,yunoki_,yunoki2} and in the previous section 
for the 1-orbital problem was used. 
Finally, to analyze orbital correlations the pseudopin operator
${\bf T}_{\bf i} = {{1}\over{2}} \sum_{{\sigma}ab} 
c^{\dagger}_{{\bf i}a \sigma} {\bf \sigma}_{ab} c_{{\bf i} b \sigma}$
was used, while for spin correlations the operator is 
standard. The Fourier-transform of the pseudospin correlations is
defined as $T({\bf q}) = {{1}\over{L}} \sum_{\bf l,m} e^{i{{\bf
q}\cdot{\bf (l-m)}} } \langle 
{ {{\bf T}_{\bf m}}\cdot{{\bf T}_{\bf l}} } \rangle $,
with a similar definition for the spin structure factor
$S({\bf q})$.

\begin{figure}[htb]
\vspace{-0.5cm}
\centerline{\psfig{figure=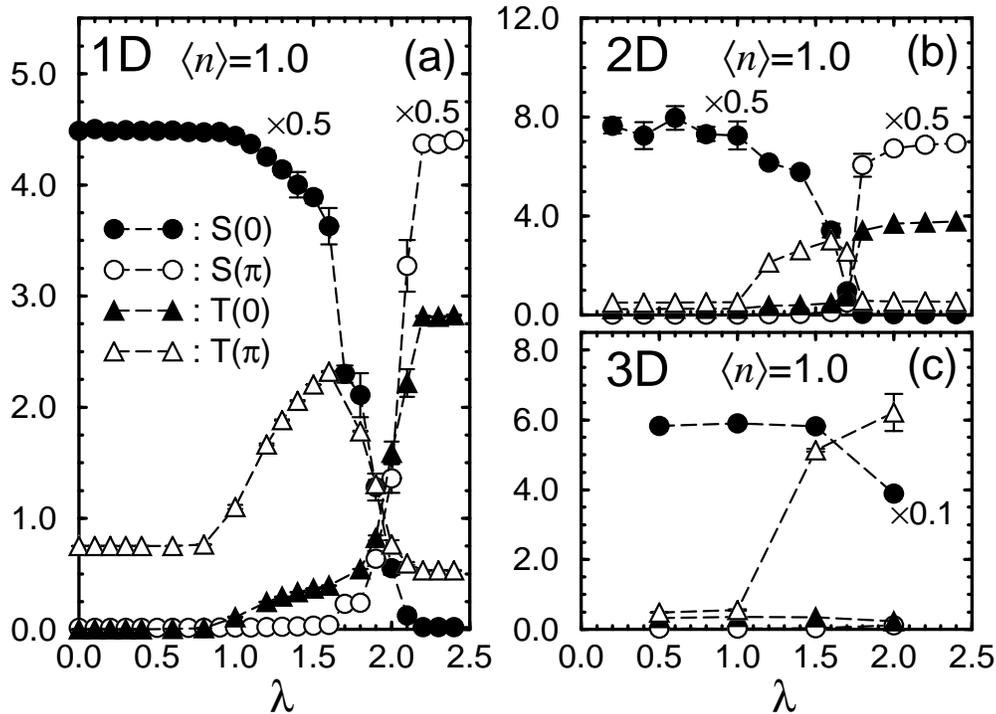,width=14.0cm,angle=-90}}
\vspace{0.2cm}
\caption{
(a) $T(q)$ and $S(q)$ vs $\lambda$, working at $\langle n \rangle =1.0$,
$T=1/75$, 
$J_H =8$, $J'=0.05$, and in 1D with $10$ sites. 
$\{ t_{ab} \}$ correspond to set $T_1$ (see [28]).
Results with sets $T_2$ and $T_3$ are qualitatively the same;
(b) Same as (a) but for a $4^2$
 cluster, $T=1/50$, and hopping $T_3$ ($T_4$) in the $y$ ($x$)
direction.
$q= 0 ( \pi )$ denotes $(0,0)$ ($(\pi,\pi)$); 
(c) Same as (a) but for a $4^3$ cluster,
$T=1/50$, the 3D 
hopping amplitudes of Ref.~[28], and $J_H = \infty$.
$q= 0 ( \pi )$ denotes $(0,0,0)$ ($(\pi,\pi,\pi)$).
These results were taken from Ref.~[11].
}
\end{figure}

Let us first consider the limit $\langle n \rangle = 1.0$,
corresponding to undoped manganites.
Fig.5 shows $T(q)$ and $S(q)$
at representative momenta $q=0$ and $\pi$
vs. $\lambda$. For small electron-phonon coupling 
the results are similar to those at $\lambda = 0.0$,
namely a large $S(0)$
indicates a tendency to  spin-FM order induced by DE (as
in the qualitatively similar 1-orbital
problem at $\langle n \rangle = 0.5$~\cite{yunoki}).
The small values of $T(q)$ imply that in this regime the
orbitals  remain
disordered.
When the coupling reaches $\lambda_{c1} \sim 1.0$, the rapid increase of
$T(\pi)$ now suggests that the ground state has a tendency to form a
 $staggered$ (or ``antiferro'' AF)
 orbital pattern, with the spins remaining FM aligned
since $S(0)$ is large.
The existence of this phase was discussed before, but
using multi-orbital Hubbard models 
with Coulomb interactions and
without phonons~\cite{orbital}. Our results show that it can
be induced by JT phonons as well. 
As the coupling increases further, another transition 
at $\lambda_{c2} \sim 2.0$ occurs 
to a spin-AF orbital-FM state ($S(\pi)$ and $T(0)$ are
large). In this region a 1-orbital
approximation is suitable.

The three
regimes of Fig.5 can be understood in the limit 
where $\lambda$ and $J_H$ are the largest scales, and using
$t_{12}=t_{21}=0$, $t_{11}=t_{22}=t$  for simplicity. For parallel
spins with orbitals split in a staggered (uniform) pattern, the energy
per site
at lowest order in $t$ is $\sim -t^2/\Delta$ ($\sim 0$), where
$\Delta$ is the orbital splitting. For antiparallel spins with
uniform (staggered) orbital splitting, the energy is $\sim -t^2/2J_H$
($\sim -t^2/(2J_H + \Delta)$). Then, when $\Delta <  2 J_H$ (``intermediate''
$\lambda$s), a spin-FM orbital-AF
order dominates, while as $\lambda$ grows further a transition to a
spin-AF orbital-FM ground state is expected.
Note that this reasoning is actually valid in any dimension.
To confirm this prediction, results
for a 2D cluster in Fig.5b are shown.
Indeed
the qualitative behavior is very similar to the 1D results.
In 3D (Fig.5c), where studies on $4^3$ clusters can only be done at
 $J_H=\infty$ to reduce the number of DOF, 
at least two of the regimes of Fig.5a have
been identified~\cite{comm66}.

Let us discuss the transport properties
in the three regimes 
at $\langle n \rangle = 1$. The algorithm used in Ref.~\cite{yunoki,yunoki3} 
allow us to
calculate {\it real-time} dynamical responses accurately, including  
the optical conductivity $\sigma(\omega > 0)$,
since all the eigenvectors in
the fermionic sector for a given spin and phonon
 configuration are
obtained exactly. From the sum-rule,
$e_g$ kinetic-energy,
and the integral of $\sigma(\omega > 0)$,
the $\omega = 0$ Drude-weight $D_W$ can be obtained.
Carrying out this calculation it was observed that 
$D_W$ vanishes at $\lambda_{c1}$
signaling a {\it metal-insulator} transition (MIT). 
Here the insulating phase is
spin-FM and orbital-AF, while
the metallic one is spin-FM and orbital-disordered~\cite{comm60}.
The density of
states (DOS) for $\lambda > \lambda_{c1}$ was also calculated 
in Ref.~\cite{yunoki3} and it
presents a clear gap at the Fermi level.
The qualitative shape of $D_W$ vs $\lambda$ on  $4^2$ and
$4^3$ clusters was found to be the same as in 1D and, thus, it is reasonable
 to assume that the
MIT exists also in all dimensions of interest.

Consider now the influence of hole doping on the $\langle n
\rangle =1.0$ phase diagram, with special emphasis on the stability of
other densities as $\mu$ is varied.
Fig.6 shows $\langle n \rangle $ vs $\mu$ in
the intermediate-$\lambda$ regime. 
It is remarkable
that $two$ regions of unstable densities exist
at low-$T$ (similar conclusions
were reached  using the Maxwell's construction~\cite{comm67}).
These instabilities 
signal the existence of PS in the $H_{KJT}$ model.
At low-density there is separation between
 an (i) empty $e_g$-electron band with AF-ordered $t_{2g}$-spins 
and a (ii) metallic spin-FM orbital uniformly ordered phase.
This regime of PS is
analogous to the spin-driven PS found at low-density in the 1-orbital 
problem~\cite{yunoki2}.
On the other hand,
the unstable region near $\langle n \rangle =1.0$ is $not$ contained in
the 1-orbital case. Here PS is between the phase
(ii) mentioned above,
and (iii) the insulating intermediate-$\lambda$ spin-FM and orbital-AF 
phase described in Fig.5~\cite{maekawa}.
In 2D the results were found to be very
similar (Ref.~\cite{yunoki3}).
The driving force
for this novel regime of PS are the {\it orbital} degree of freedom, since
the spins are uniformly ordered in both phases involved.
Studying $\langle n \rangle$ 
vs $\mu$, for $\lambda < \lambda_{c1}$
only PS at small densities is observed, while
for $\lambda > \lambda_{c2}$ the PS close to $\langle n \rangle = 1$
is similar to the same phenomenon observed in the 1-orbital problem since
it involves  a spin-AF orbital-FM phase~\cite{yunoki}.

\begin{figure}[htb]
\vspace{-0.5cm}
\centerline{\psfig{figure=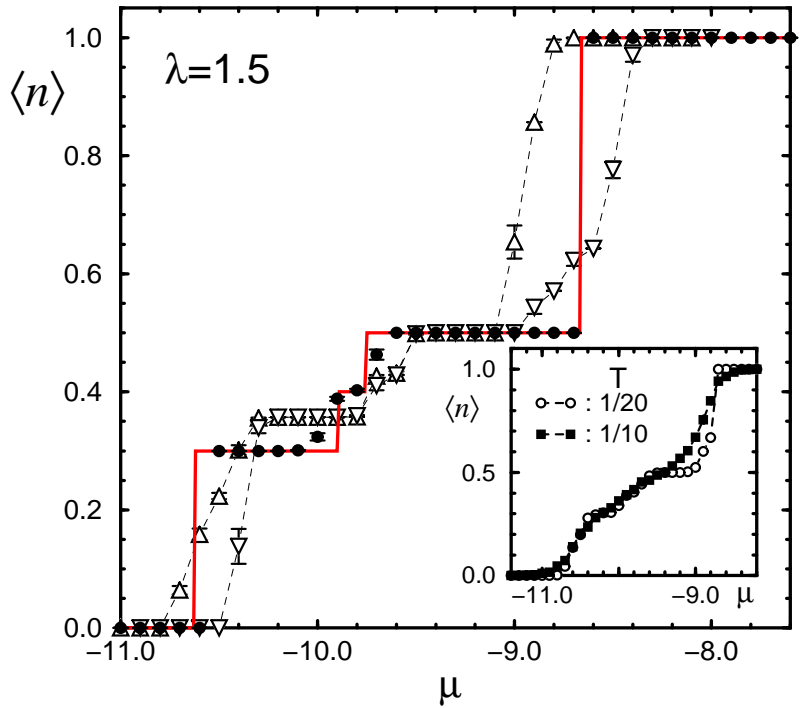,width=14.0cm,angle=-90}}
\vspace{0.2cm}
\caption{
$\langle n \rangle$ vs $\mu$ at $\lambda = 1.5$, $L=10$, and
$T=1/40$ (solid circles).
The discontinuities near $\langle n \rangle =1.0$
and $0.0$ show the existence of unstable densities. 
The solid line is
obtained from the Maxwell's construction. The triangles are results also at
$\lambda =1.5$ and $T=1/40$, but using 14 sites and only $2 \times 10^4$ MC
sweeps to show the appearance of {\it hysteresis} loops
as in a first-order transition.
The inset shows the
$T$-dependence of the results at $L=10$.
These results were taken from Ref.~[11].
}
\end{figure}

The phase diagram of the 1D $H_{KJT}$
model is given
in Fig.7. The two PS regimes are shown, together with the
metallic spin-FM region. This phase
is separated into two regions, one ferro-orbital ordered and
the other orbitally disordered. The existence of these two
regimes can be deduced from the behavior of the pseudospin
correlations, the mean value of the pseudospin operators,
and the probability of double occupancy of the same site
with different orbitals.
The results are similar for several
$\{ t_{ab} \}$ sets~\cite{hopping}. 
The simulations of Ref.~\cite{yunoki3} suggest that the qualitative
shape of Fig.7 should be valid also in $D=2$ and $3$.
Recently we learned of experiments where the 
low-$T$ coexistence of domains similar to those described in 
Ref.~\cite{yunoki3}, i.e.
one orbitally-ordered and the other FM-metallic,  has been
observed in $\rm Mn$-perovskites~\cite{radaelli,kubo}.

\begin{figure}[htb]
\vspace{-0.5cm}
\centerline{\psfig{figure=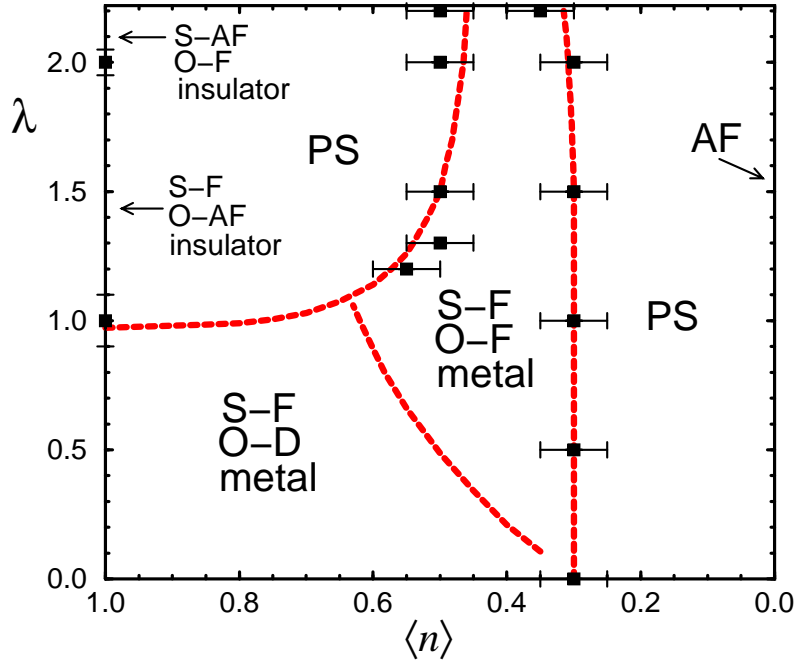,width=14.0cm,angle=-90}}
\vspace{0.2cm}
\caption{
Phase diagram of the $H_{KJT}$ model at low-$T$, $J_H=8.0$, $J'=0.05$,
and using set $T_1$ for  $\{ t_{ab} \}$.
$\rm S-F$ and $\rm S-AF$ denote regimes with FM- and AF-spin 
orders, respectively. $\rm O-D$, $\rm O-F$, and $\rm O-AF$ 
represent states with 
disordered, uniform and staggered orbital orders, respectively.
PS means phase separation.
These results were taken from Ref.~[11].
}
\end{figure}

\begin{flushleft}
{\bf Optical Conductivity}
\end{flushleft}

In Ref.~\cite{yunoki3} results for  $\sigma(\omega)$ were also presented.
Experimental studies for a variety of manganites
such as ${\rm Nd}_{0.7} {\rm Sr}_{0.3} {\rm Mn} {\rm O}_3$, 
${\rm La}_{0.7} {\rm Ca}_{0.3} {\rm Mn} {\rm O}_3$, and 
${\rm La}_{0.7} {\rm Sr}_{0.3} {\rm Mn} {\rm O}_3$ reported
a broad peak at $\omega \sim 1eV$ (for hole doping $x > 0.2$ and 
$T > T^{FM}_c$)~\cite{kaplan,jung}.
At room-$T$ 
there is negligible weight near $\omega=0$,
but as $T$ is reduced the $1eV$-peak shifts to smaller energies, gradually
transforming into a Drude response well below $T_c$. The
finite-$\omega$
peak can be identified even inside the FM
phase. The coherent spectral weight is only a small fraction
of the total. Other features at larger energies $\sim 3eV$
involve transitions between the $J_H$-split bands and the $O$-ions.
In addition, Jung et al.~\cite{jung}
interpreted the $1eV$ feature at room-$T$ as composed of two
peaks due to intra- and inter-atomic transitions in JT-distorted 
environments.

In Fig.8, $\sigma(\omega)$ for the $H_{KJT}$ model is shown 
at $\langle n \rangle = 0.7$ and several temperatures 
near the unstable PS
 region of Fig.7 (weight due to $J_H$
split bands occurs at higher energy).
Here the FM spin correlation length 
grows rapidly with the lattice size for $T^* \leq 0.05t$, which can be
considered as the ``critical'' temperature. 
Both at high- and intermediate-$T$ a broad
peak is observed at $\omega \sim 1$, smoothly evolving to
lower energies as $T$ decreases. The peak can be identified well-below
$T^*$ as in 
experiments~\cite{kaplan,jung}. 
Eventually at very low-$T$, $\sigma(\omega)$ is
dominated by a Drude peak.
The $T$-dependence shown in
the figure is achieved at this $\lambda$ and $\langle n \rangle$ 
by a combination of a finite-$\omega$ phonon-induced broad feature
that looses weight, and a
Drude response that grows as $T$ decreases (for smaller $\lambda$s, 
the two peaks can be distinguished even at low-$T$).
The similarity with
experiments suggests that real
manganites may have couplings close to an unstable 
region in parameter space.
In the inset, $D_W$
vs $T$ is shown. Note that
at $T \sim T^*$,
$D_W$ vanishes suggesting a MIT, probably
 due to magneto polaron localization. Results for the 1-orbital case
are smoother, with no indications of a singularity.
These features are in agreement with experiments, since the
manganite ``normal''
state is an insulator. Work is currently in progress to analyze in
more detail this MIT transition.

\begin{figure}[htb] 
\vspace{-0.5cm}
\centerline{\psfig{figure=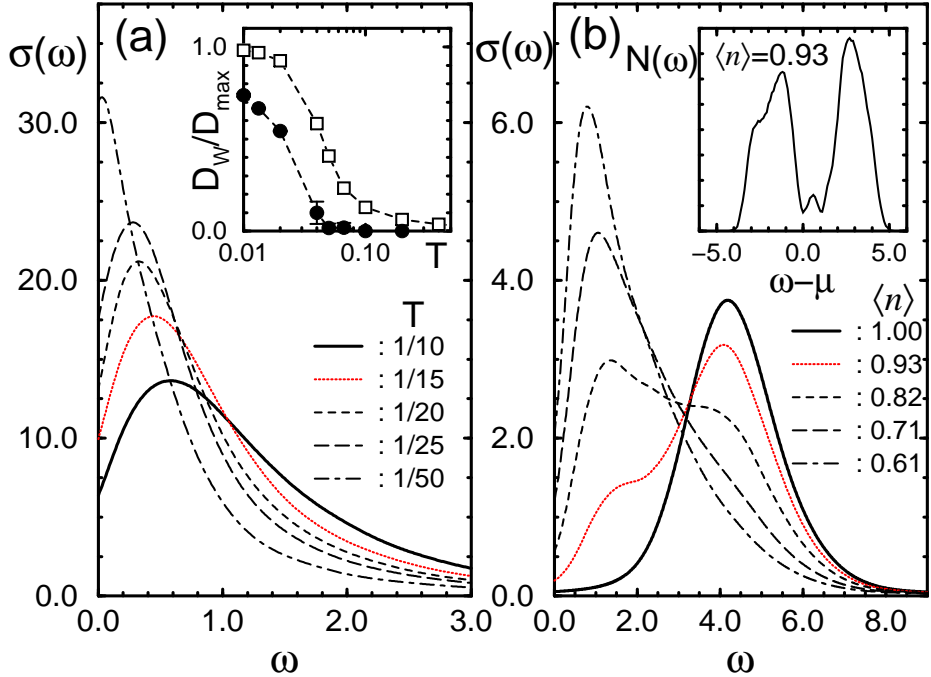,width=13.0cm,angle=-90}}
\vspace{0.2cm}
\caption{
(a) $\sigma(\omega)$ parametric with $T$,
at $\lambda=1.0$, $\langle n \rangle = 0.7$, and $L=20$.
The inset shows $D_W$ vs $T$ for both the $H_{KJT}$ (soild circles) 
and the 1-orbital model of Ref.~[8] (open squares) 
(the latter at $\langle n \rangle = 0.65$). $D_W$ is
normalized to its maximum value at $T=0.01$; 
(b) $\sigma(\omega)$ vs $\omega$ parametric with
$\langle n \rangle$ at $\lambda = 1.5$, $T=1/10$, and $L=16$ (results
for $L=10$ are very similar).
The inset shows the lower $J_H$-split
DOS at $\langle n \rangle = 0.93$.
Both in (a) and (b) a
$\delta$-function broadening $\epsilon = 0.25$ was used.
These results were taken from Ref.~[11].
}
\end{figure}

A similar good agreement with experiments was observed working
in the regime of the orbitally-induced PS but at a temperature above
its critical value $T_{PS}$ (roughly 
$\sim 1/20$, see inset Fig.6). Here 
 the broad feature observed at high-$T$ in Fig.8a moves to 
higher energies (Fig.8b) since $\lambda$ has increased.
At the temperature of
the plot the system is an insulator at $\langle n \rangle =
1$, but as hole carriers are added a second peak at lower energies
develops, in addition to a weak Drude peak (which carries, e.g.,
 just $1\%$ of the total weight at $\langle n \rangle = 0.61$). This 
feature at high-$T$ is reminiscent of recent experimental
 results~\cite{jung} where a two-peak 
structure was observed at room-$T$ and several
 densities. Similar results were obtained on $4^2$
clusters. In Fig.8b the peak at large-$\omega$
 is caused by
phononic effects since its position was found to grow 
rapidly with $\lambda$.
It corresponds to intersite transitions between $\rm Mn^{3+}$ JT-split states.
The lower energy structure is compatible with a $\rm Mn^{3+}$-$\rm Mn^{4+}$
transition~\cite{warning}. The inset of Fig.8b shows the DOS
 of the system.
The two peaks above $\mu$ are responsible for the features found
in $\sigma(\omega)$. This interpretation is the same
as given in Ref.~\cite{millis2} at $D=\infty$.

\begin{flushleft}
\vspace{24pt}
{\bf EXPERIMENTAL CONSEQUENCES OF PHASE SEPARATION}
\end{flushleft}

A large number of papers in the context of experimental studies
of manganites have reported the presence of some sort of inhomogeneity
in the system that is tempting to associate with the phase-separation
tendencies observed in our studies. In the next paragraph we will
mention some of these results such that the interested reader
can have at hand at least part of the relevant references on the subject to
decide by him/herself on this issue.

Some of the experiments that have reported  results that could be
compatible with ours are the following (in a random order): {\bf 1.}
De Teresa et al.~\cite{teresa} studied $\rm La_{1-x} Ca_x Mn O_3$ 
using small-angle neutron scattering,
magnetic susceptibility and other techniques, at x=1/3. The analysis of
their data showed the existence of ``magnetic clusters'' of size
approximately $12 \AA$ above the ferromagnetic critical temperature;
{\bf 2.} Hennion et al.~\cite{hennion} recently presented elastic 
neutron scattering results
below $T_c$ at $\rm x=0.05$ and $\rm 0.08$ also for $\rm La_{1-x} Ca_x Mn O_3$.
They interpreted their results as indicative of ``magnetic droplets''. The
density of these droplets was found to be much smaller than the density of
holes implying that each droplet contains several holes; {\bf 3.} Lynn and
collaborators~\cite{lynn} have studied  $\rm La_{1-x} Ca_x Mn O_3$ at x=1/3
also using neutron scattering. Lattice anomalies and magnetic irreversibilities
 near $T_c$ were interpreted as evidence of two coexisting distinct phases;
{\bf 4.} Perring et al.~\cite{perring} studying 
$\rm La_{1.2} Sr_{1.8} Mn_2 O_7$ with
neutron scattering reported the presence of long-lived antiferromagnetic
clusters coexisting with ferromagnetic critical fluctuations above $T_c$;
{\bf 5.} Allodi et al.~\cite{allodi} using NMR applied to
$\rm La_{1-x} Ca_x Mn O_3$ at ${\rm x}\sim 0.1$ observed coexisting 
resonances corresponding to FM and AF domains; 
{\bf 6.} Cox et al.~\cite{cox} studying
$\rm Pr_{0.7} Ca_{0.3} Mn O_3$ with x-ray and powder neutron scattering
reported the presence of ferromagnetic clusters; 
{\bf 7.} Bao et al.~\cite{bao}
in their analysis of $\rm Sr_{2-x} La_x Mn O_4$ (2D material) found
phase separation at small $e_g$-densities; 
{\bf 8.} Yamada et al.~\cite{yamada}
in their study of $\rm La_{1-x} Sr_x Mn O_3$ with neutron scattering at
x=0.1 and 0.15 interpreted their results as corresponding to
polaron ordering; {\bf 9.} Roy et al.~\cite{roy} studying
$\rm La_{1-x} Ca_x Mn O_3$ near $x=0.50$ found the coexistence of two
carrier types: nearly localized carries in the charge-ordered state and
a parasitic population of free carriers tunable by stoichiometry;
{\bf 10.} Booth et al.~\cite{booth} also studying $\rm La_{1-x} Ca_x Mn O_3$
found that the number of delocalized holes $n_{dh}$ in the ferromagnetic phase
changes with the magnetization $M$ as $ln(n_{dh}) \propto -M$.
Since the magnetization saturates well below $T_c$, this
implies that there is a range of temperatures where sizable fractions
of localized and delocalized holes coexist;
{\bf 11.} Jaime et al.~\cite{jaime} reported the possibility of polaronic
distortions of the paramagnetic phase of $\rm La-Ca-Mn-O$ manganites
persisting into the ferromagnetic phase, and analyzed the data with
a two-fluid model; {\bf 12.} Zhou and Goodenough~\cite{zhou} studying the
thermopower and resistivity of $\rm ^{16}O/^{18}O$ isotope-exchanged
$\rm (La_{1-x} Nd_x )_{0.7} Ca_{0.3} Mn O_3$ found indications of a
segregation of hole-rich clusters within a hole-poor matrix in the
paramagnetic state; {\bf 13.} Billinge et al.~\cite{billinge} reported the
coexistence of localized and delocalized carriers in
a wide range of densities and temperatures below $T_c$; 
{\bf 14.} Ibarra et al.~\cite{ibarra} studied 
$\rm (La_{0.5} Nd_{0.5} )_{2/3} Ca_{1/3} Mn O_3$ concluding that
insulating charge-ordered and metallic ferromagnetic regions coexist at
low temperatures; {\bf 15.} Rhyne et al.~\cite{rhyne} in their study of 
$\rm La_{0.53}Ca_{0.47} Mn O_3$ with neutron diffraction found two
magnetic phases below the transition temperature, one
ferromagnetic and the second antiferromagnetic, both
persisting down to 10K; {\bf 16.} Heffner et al.~\cite{heffner} studying
$\rm La_{0.67} Ca_{0.33} Mn O_3$ using zero-field muon spin relaxation 
and resistivity techniques found indications of polarons on the spin 
and charge dynamics; {\bf 17.} Jung et al.~\cite{jung2} have very 
recently studied the optical conductivities of 
$\rm La_{7/8} Sr_{1/8} Mn O_3$ concluding that there are indications in 
this system of phase separation.

The common theme of these experiments, and possibly others that have
escaped our attention, is that some sort of inhomogeneity appears
in the analysis of the data. The phenomenon is apparently more clear
at small hole densities x and low temperatures which is precisely the
region where our results indicate a strong tendency to phase separate.
But also above $T_c$ in the interesting densities around x=1/3
magnetic clusters have been reported. It is nothing but natural to
imagine a ``smooth'' connection between ${\rm x}\sim 0.1$ and low temperature
with ${\rm x}\sim 1/3$ and temperatures above $T_c$. In this case the large
droplets found by Hennion et al.~\cite{hennion} could have evolved as 
to become the magnetic clusters of De Teresa et al.~\cite{teresa} 
by a reduction of their size and
number of holes inside. We strongly encourage experimental work
that could contribute to the answer of this or other  important issues
around the common theme of possible charge seggregation in real manganites.

\begin{flushleft}
\vspace{24pt}
{\bf SUMMARY}
\end{flushleft}

Summarizing, a comprehensive computational study of models for manganites
have found that the expected double-exchange induced
strong tendencies to ferromagnetic correlations
at low temperatures
are in competition with  a regime of
``phase separation''. This regime was identified
in all dimensions of interest, using one and two orbitals (the latter with
Jahn-Teller phonons), and both with classical and quantum localized
$\rm t_{2g}$ spins. It also appears in the presence of on-site Coulomb
interactions. This robustness of our results suggests  that phase separation
may also be present in real manganites. In the previous section
experimental literature that have reported some form of charge inhomogeneity
in the context of the manganites has been briefly reviewed. It is concluded
that theory and experiments seem to be in qualitative agreement and
phase separation tendencies (which may correspond to the formation of
magnetic droplets or even stripes once Coulomb interactions beyond the
on-site term are included in the analysis) should be taken seriously.
They may even be responsible for the phenomenon of Colossal
Magnetoresistance that motivated the current enormous interest in the
study of manganites in the first place!.

\begin{flushleft}
\vspace{24pt}
{\bf ACKNOWLEDGEMENTS}
\end{flushleft}

E. D. and A. M. are supported by the NSF grant DMR-9520776. 
S. Y. thanks the NHMFL for support.


\begin{thebibliography}{99}
\bibitem{foot3} In both 1D and 2D the Mermin-Wagner theorem forbids
a nonzero $T_c$ in the model. However, the
correlation lengths can be very large even at finite $T/t$.
\bibitem{approx} This approximation was shown to be
accurate in Ref.~\cite{yunoki,yunoki_}.
\bibitem{comm66}  Studies at
larger values of $\lambda$'s are 
too costly in CPU time.
\bibitem{comm60} In Ref.~\cite{millis2}, a MIT at $\lambda \sim 1$ was
also found but it was not associated with orbital order.
\bibitem{comm67} The
energies needed for this construction are obtained
at low-$T$ from the MC evolution of the density~\cite{yunoki2}.
\bibitem{warning} Although its dependence with $\lambda$ was weak.
\end{thebibliography}
\end{document}